\documentclass[journal]{IEEEtran}

\usepackage[dvips]{graphicx}

\graphicspath{{figures/}}
\DeclareGraphicsExtensions{.eps}

\usepackage[cmex10]{amsmath}

\usepackage{amssymb}

\usepackage{cite}

\begin{document}

\title{Probing the sheath electric field using thermophoresis in dusty plasma. \\Part II: Experimental measurements}

\author{Victor~Land, Bernard~Smith, Lorin~Matthews, Truell~Hyde~\IEEEmembership{Fellow,~IEEE}
\thanks{V. Land, B. Smith, L. Matthews, and T. Hyde are with the Center for Astrophysics, Space Physics and Engineering
Research, at Baylor University, Waco, TX, 76798-7316 USA, e-mail: victor\_land@baylor.edu,  (see http://www.baylor.edu/CASPER).}
\thanks{Manuscript received xxxxx xx, 2009; revised xxxxx xx, 2009.}}

\markboth{IEEE transactions on plasma science,~Vol.~XX, No.~XX, XXXXX~2010}%
{Land \MakeLowercase{\textit{et al.}}: Dust particles as probes for the sheath electric field by applying additional
thermophoresis}

\maketitle

\begin{abstract}
A two-dimensional dust crystal levitated in the sheath of a modified Gaseous Electronics Conference (GEC) reference cell is manipulated
by heating or cooling the lower electrode. The dust charge is obtained from top-view pictures of the crystal using a
previously developed analytical model. By assuming a simple force balance, and measuring the radial confining force, the
vertical electric field profile in the sheath is reconstructed. The dust crystal is shown to levitate on
the plasma side of the Bohm point. Finally, it is shown that the ion drag plays an important role in the vertical force
balance, even for large dust grains.
\end{abstract}


\IEEEpeerreviewmaketitle

\section{Introduction}

\IEEEPARstart{D}{ust} particles immersed in laboratory plasma collect electrons and ions and become negatively charged. The
charge on a single particle is
typically estimated to be between 1 and 5 electron charges per nanometer radius \cite{Stoffels1999}, depending on the background
pressure \cite{Ratynskaia2004}. This means that micrometer-sized particles used in dusty plasma experiments can carry a negative charge of
thousands of electron charges. Because of this, the dust particles are levitated against gravity in the strong electric
fields present in the sheath. The electrostatic interaction between the particles is shielded by the plasma, but depending on
the discharge parameters, strongly coupled systems can be formed \cite{Thomas1994}. By applying a radial confinement, these so
called dust crystals can be used to study many solid state physics phenomena on a scale accessible to ordinary optical
diagnostic techniques.

In order to understand the behavior of these systems, knowledge about the screening length and the dust charge is of paramount importance. Furthermore, once the charge on the dust particles is known, the
vertical force balance between gravity and the electric field can provide insight into the structure of the electric field in
the sheath. This is important for both dusty plasma experiments and many plasma enhanced industrial processes, where
ions can be accelerated by the sheath electric field from the plasma bulk to a processing substrate \cite{Liebermanbook}.

A method for determining the Debye length and dust particle charge for a dust crystal suspended above a spherical depression in the lower
electrode of a RF discharge, by measuring observable quantities in top-view pictures of dust crystals, was
developed in \cite{Hebner2002}. This method was recently investigated computationally and the results presented elsewhere in this
issue \cite{partI}. 

In this paper, we present experimental results of the determination of the Debye length and the dust
charge for two-dimensional dust crystals levitated in a modified GEC cell. 
By using thermophoresis to move the dust crystal
through the sheath, the electric field was reconstructed based on the measured dust charge and the assumed
balance between gravity and the electrostatic force in the sheath. It is shown that this method is consistent with our
computational results, that the ion drag most likely plays an important role in the vertical force balance, and finally that
the dust crystal is levitated on the plasma side of the classical Bohm point.

\section{Experimental setup}\label{sec:setup}

A GEC cell was used \cite{Hargis1994} to create a standard 13.56 MHz
radio-frequency (RF) discharge in argon. The cell was modified in order to conduct dusty plasma experiments. Cover plates can
be attached on top of the powered bottom electrode, which have cylindrically shaped depressions milled in them, with
different diameters. These \emph{cutouts} provide an electric field that confines the
dust particles radially. The upper electrode is a hollow grounded cylinder, which provides optical access to the dust
crystal. Above this electrode dust containers are installed, which, when tapped, release dust particles which fall through
the hollow electrode into the plasma, until they reach their equilibrium levitation height.

A camera mounted above the
hollow upper electrode is used to provide top-view pictures of the horizontal dust crystal structure. A second camera installed in front of a
side-port takes side-view pictures of the vertical structure of the dust crystal. Two red diode-lasers with a wavelength of
640 nm and 686 nm are equipped with cylindrical lenses
that shape the laser beams into thin laser sheaths. One sheath is oriented vertically, the other horizontally. The light
scattered by the dust particles is captured by either the top-view or side-view camera. Filters can be used with the
cameras to only allow light at the laser wavelength to be observed, rejecting the plasma glow. Typically, only one camera-laser
combination was used at a time. The cameras and lasers can be moved in three directions; however, the focus of
the cameras is
fixed, so that the resolution did not change between pictures. The resolution of the top-camera and side-camera was
measured to be 26 and 14 $\mu$m per pixel, respectively.

\begin{figure}[!ht]
\centering
\includegraphics[width=2.5in]{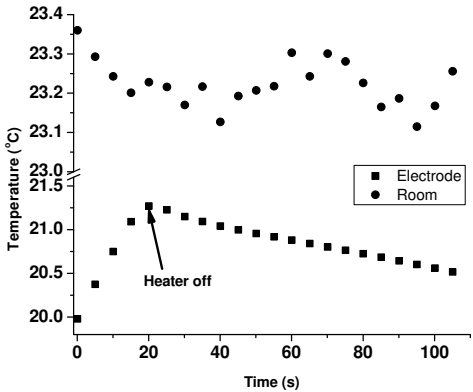}
\caption{The room-temperature and the temperature of the powered electrode measured with thermocouples. The
room-temperature was roughly constant at 23${}^{\circ}$C. The temperature on the electrode was set at 20${}^{\circ}$ C. It can be seen that the heater/thermostat rapidly heats the system, but then overshoots by a
few degrees. After reaching the maximum temperature, the electrode slowly cools down, giving a relatively long time-window
where the temperature may be considered fairly constant.}
\label{fig:temperaturechange}
\end{figure}

The lower electrode is constructed such that it can be cooled or heated by running liquid through tubes inserted in a circular channel drilled
in the electrode. For heating and cooling the liquid, we used a MGW Lauda Brinkman RM3T chiller/heater, connected to a
Ranco ETC-111000-000 thermostat. The temperature was measured employing thermocouples attached to the tube near the thermostat, as well as on the
bottom of the electrode. Due to the hysteresis of the thermostat/chiller-heater system, the temperature
fluctuated approximately 2${}^{\circ}$ C above and below the set temperature. A rapid rise to temperatures above
the set temperature was followed by a slow decrease to a point 2 degrees below the set-point. By measuring the electrode-temperature
variation, we determined an interval of at least 60 seconds where the temperature change was slow and the temperature could
be considered constant. This was long enough to take the necessary
top- and side-view pictures, since the cameras operated at 60 and 30 Hz, respectively. Figure \ref{fig:temperaturechange}
shows the change in temperature of the lower electrode measured by a thermocouple attached to the bottom of the electrode,
together with the ambient temperature. To determine the electric field
profile, the set temperature was varied between 0${}^{\circ}$ C and 60${}^{\circ}$ C in steps of 10${}^{\circ}$ C.

For the experiments presented here, 8.89 $\mu$m diameter melamine-formaldehyde (MF) particles were used, in an argon
discharge run at 2 W of input power and 200 mTorr of background pressure. 
The diameter of the cutout in the top-plate attached to the lower electrode was 1 inch (25.4 mm) with the depth of the cutout
roughly half a millimeter.

\section{Observables}\label{sec:Observables}

A complete description of the analytical method used to determine the Debye length and dust charge was developed in \cite{Hebner2002} and
discussed in \cite{partI}. We here restrict ourselves to mentioning only the observables required to determine the Debye length and
the dust charge, together with the equations relating these to the observables.

Given a single-layer dust crystal consisting of $N$ particles, the radius of the crystal $R_M$, the inter-particle
spacing at the edge of the crystal, ${\Delta}_M$, and the inter-particle spacing at the center of the crystal ${\Delta}_0$
were measured. The theoretical edge of the crystal, $R_{\infty} = R_M + {\Delta}_M\sqrt{3/2}$ was then calculated, so that the
Debye length could be determined from

\begin{equation}\label{eq:lambda}
{\lambda}_D = {\Delta}_0\left[\frac{A-S}{3S-A}\right],
\end{equation}

\noindent where we have defined the total crystal surface area as $A = \pi R_{\infty}^2$ and the total surface area covered by $N$
Wigner-Seitz cells measured at the center to be $S = \sqrt{3}{\Delta}_0^2N/2$.

After determining the Debye length, the dust charge can be calculated from 

\begin{equation}\label{eq:charge}
q_D = \sqrt{\frac{4\pi{\epsilon}_0{\Delta}_0kR_{\infty}^2}{3\left(3+\frac{3S-A}{A-S}\right)\exp(\frac{A-3S}{A-S})}},
\end{equation}

\noindent where $k$ is the constant of the radial restoring force. It is assumed that the radial potential well is parabolic, so
that $V(r) \propto cr^2$. In this case, $k$ depends on the steepness of the potential well, which depends on the
geometry of the cutout.
Also, $k$ changes with the applied temperature, and hence with any change in the thermophoretic force $F_{th}$; $k=2c\left[m_Dg
- F_{th}\right]$. 
We see that when the thermophoretic force acts downward ($F_{th} < 0$), $k$ increases and the dust crystal is radially compressed. When
the thermophoretic force acts upward, $k$ decreases and the dust crystal expands radially. 

In \cite{Hebner2002} spherical cutouts were used, and the curvature of the potential well, $c$, was related to the radius
of curvature of the cutout, $R_C$; $c=1/(2R_C)$. We will assume a similar relation for our cylindrical cutout, so that we have $k=\left[m_Dg
- F_{th}\right]/R_C$.

\subsection{Measuring $k$ without thermophoresis}

The common assumption concerning the radial confinement resulting from cutouts or other adaptations to the lower electrode is that
it can be described by a potential well varying quadratically with the distance from the center. Even though this has been shown
for specific experiments \cite{Konopka2000}, this is not \textit{a priori} clear for our case.

\begin{figure}[!ht]
\centering
\includegraphics[width=2.5in]{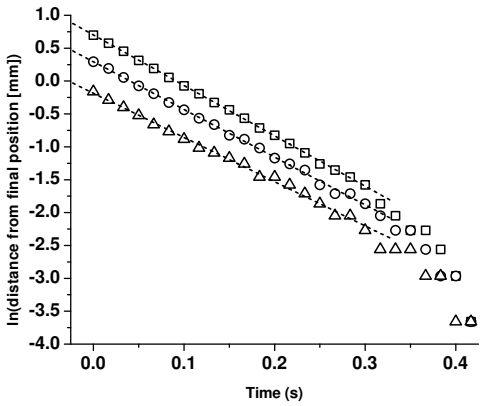}
\caption{Three selected particle trajectories observed with the top-mounted camera at 1000
mTorr. Plotted are the natural logarithm of the distance from the final position in the last frame. For the first 0.3
seconds, the trajectories form straight lines. The
dashed lines are linear fits to the natural logarithm of the distance, all with $R^2>0.996$. From these trajectories the value of $k_0$ is determined.}
\label{damped_motion}
\end{figure}

In order to determine the radial restoring force, we ran the discharge at high pressure (1 Torr in this case) in order to
minimize any oscillatory motion of the dust particles, without heating or cooling of the lower elecotrode, so that $F_{th} = 0$, and dropped the 
dust particles just outside of the cutout. We then monitored the particles while they moved inwards into the potential well of
the cutout. Assuming that in this case the forces acting on the particles are the electrostatic force (determined by the potential well)
and the gas drag, the equation of motion for these particles can be written as:

\begin{equation}\label{eq:overdamped}
m_D\ddot{x} = -\alpha\dot{x} - k_0 x,
\end{equation}

\noindent where the first term on the right hand side represents the neutral drag, with $\alpha$ the Epstein friction
coefficient \cite{Epstein}, and the
second term the restoring force. Note that here we presume the force is due to a harmonic potential, hence the use of
Hooke's law. We assume that the particles are monodisperse, so that they
all have the same mass, $m_D$. A solution to equation \ref{eq:overdamped} can be found by substituting $x(t) = a\exp(-bt)$ for
overdamped motion (for high neutral drag at 1 Torr), which gives us the value of $b$ directly: $b=-\dot{x}/x = (k_0/\alpha)$. 

Figure \ref{damped_motion} shows three selected particle trajectories. Plotted is the natural logarithm of the distance from
the final position in the last frame. During the first 0.3
seconds, the trajectories remain linear in this plot. The very last part of the trjactories no longer follows these straight
lines. This might be due to interactions with particles already present near the equilibrium position above the cutout,
changing the velocity of the incoming particles towards the end of their trajectories. The dashed lines are fits according to the overdamped solution to
equation \ref{eq:overdamped} for the first 0.3 seconds. The fits are very good ($R^2>0.996$), which shows that in this case no higher order terms play an important
role in determining the radial potential well. From the fits, we
determine $k_0$ as $k_0 = (6.68 \pm 0.296)\times m_D
\times \alpha = 2.93 \times {10}^{-10} \pm 1.3 \times {10}^{-11}$. The radius of curvature for the cutout then becomes; $R_C
= m_Dg/k_0 = 1.9$ cm, which is just larger than the physical radius of the cutout. This seems to be small when compared to radii found in
\cite{Hebner2002}, where similar crystal structures were levitated above spherical cutouts.

\section{Determination of the Debye length, charge, and electric field}

Now that we have determined $k_0$ for the 1-inch diameter cutout in our
modified GEC cell, we can take pictures of the dust crystals formed with
different applied temperature gradients. From side-view pictures the levitation
height was determined, whereas using graphics software \cite{ImageJ}, the number of
particles, crystal radius and central inter-particle distance were determined
from top-view pictures.

\begin{figure}[!ht]
\centering
\includegraphics[width=2.5in]{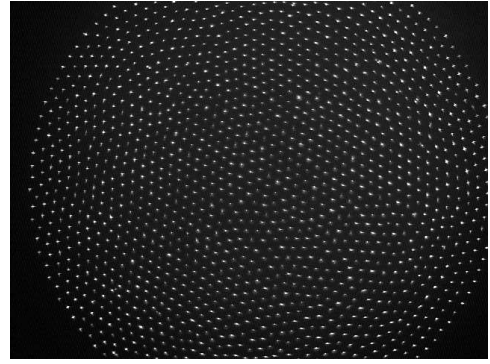}
\caption{A top-view of a dust crystal composed of 8.9 micron diameter particles, at an electrode temperature of
0${}^{\circ}$ C, applied power of 2 W, and a pressure of 200 mTorr. This
crystal has a radius of 8.2 mm, with about 1000 particles in the first layer
and a central inter-particle distance of 386 $\mu$m.}
\label{crystal_topview}
\end{figure}

Figure \ref{crystal_topview} shows a top-view image of a typical crystal levitated in the discharge. This crystal was formed
while the lower electrode was cooled to 0${}^{\circ}$ C. It contains roughly N=1000
particles in the first layer, has a radius of $R_{\infty}$= 8.2 mm, and the central
inter-particle distance was measured to be ${\Delta}_0$= 386 micron. This yields
$A=2.41 \times {10}^{-4}$ m${}^{2}$ and
$S=1.29 \times {10}^{-4}$ m${}^{2}$, so that from equation \ref{eq:lambda}, we
obtain ${\lambda}_D \approx 390 \mu$m.

\begin{figure}[!ht]
\centering
\includegraphics[width=2.5in]{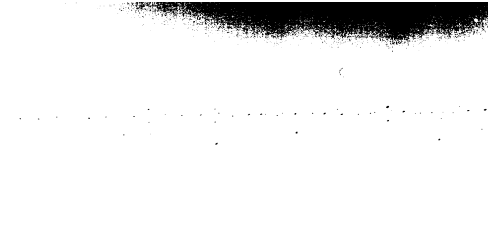}
\caption{A side-view of a dust crystal of 8.9 micron diameter particles. It can be seen that the crystal levitated in the
sheath is not completely single-layer. A slight curvature in the crystal layer is also visible, which might indicate a curvature
of the potential well. The plasma glow is visible as the dark area in the top of the picture. The colors in the picture are
inverted, the contrast of the picture has been increased with imaging software, and the image has been cropped.}
\label{fig:side-view}
\end{figure}

Since the thermophoretic force acts in the same direction as gravity when the electrode is cooled, \textit{k}
increases with decreasing lower electrode temperature. For the applied temperature gradient of 906 K/m, we find that \textit
k (0${}^{\circ}$ C) is 4.35$\times {10}^{-10}$. Using this and the value for
${\lambda}_D$ in equation \ref{eq:charge}, we find a dust charge of $q_D \approx$ -1.2$\times
{10}^{5}$ e. Using the vertical force balance, this results in an electric field of $E = (m_Dg - F_{th})/q_D \approx$ 313
V/m. From side-view
pictures, one example of which is shown in figure \ref{fig:side-view}, the levitation height of the upper 
crystal-layer was determined to be  5.83 mm. 

\begin{figure}[!ht]
\centering
\includegraphics[width=2.5in]{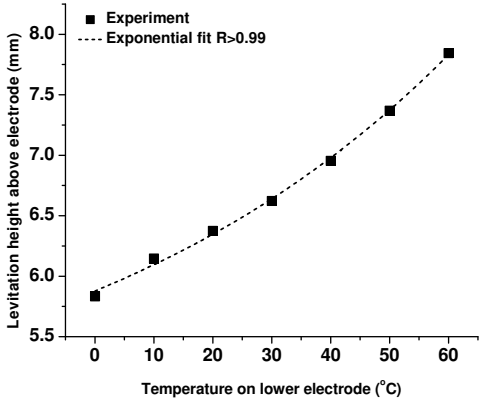}
\caption{The levitation height of the dust crystal versus temperature of the lower electrode, determined from data collected
via side-view images. An exponential fit is indicated, which shows excellent agreement with both experiment and theoretical
results \cite{partI}.}
\label{fig:lev_height}
\end{figure}

\begin{figure}[!ht]
\centering
\includegraphics[width=2.5in]{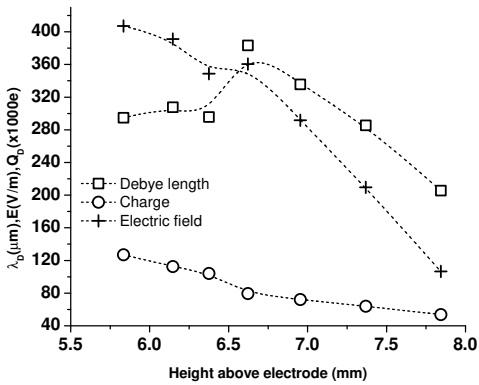}
\caption{The Debye length, dust charge, and electric field plotted versus height above
the electrode, determined from top-view and side-view pictures. The electrode
temperature was varied between 0${}^{\circ}$ C and 60${}^{\circ}$ in steps of 10
degrees. The dashed lines are there to guide the eye.}
\label{exp_results}
\end{figure}

Repeating these calculations for different
applied lower electrode temperatures, the Debye length, dust charge, and electric
field were obtained as a function of the temperature. A plot of the dust levitation height versus applied
temperature, as determined from side-view pictures, is shown in figure \ref{fig:lev_height}. It can be well fitted to an exponential, which is
consistent with the dependence of the levitation height on temperature as shown in \cite{partI}. The Debye length, dust charge, and electric field are shown in figure \ref{exp_results}.

As shown, the Debye length varies between 200 and 360 microns, having its smallest values higher in the plasma. This seems reasonable,
since the plasma density is expected to increase farther away from the electrode, and ${\lambda}_D \propto \sqrt{T/n}$. The
values obtained for the Debye length are in
reasonable agreement with values reported in the same GEC cell (roughly 250 $\mu$m for a discharge at 200 mTorr and 5 W
power), measured using a different technique \cite{Jay}, and also with 
the observed inter-particle distances, which in these crystals typically are a few times the Debye
length \cite{Thomas1994}.
 
The dust charge becomes continuously less
negative from -120,000e to -50,000e with increasing height above the lower electrode. These charges seem to be too large by a
factor of two to three, since 9-micron diameter particles are expected to have a charge of less than 50,000 electron charges. 

The electric field decreases approximately linearly for heights 
above 6.5 mm. The obtained vertical electric field in the sheath is low (due to the high reconstructed dust charge), and varies from -400 V/m
(closer to the electrode) to -120 V/m (farther up in the plasma).

However, even if we multiply the electric field by three, the drift velocity of
argon ions would correspond to an energy of roughly 0.3 eV. This is much lower
than the electron temperature, which is roughly 3-5 eV, so that the dust
is levitated on the plasma side of the Bohm point, consistent with
\cite{Hebner2002}.

\section{Discussion and conclusions}

Our results show that the application of thermophoresis provides an interesting tool to probe electric field structures in
the sheath. Debye lengths obtained in this manner appear to be in agreement with values obtained by others. The
dust charge appears to be overestimated however, roughly by a factor of two to three, resulting in unrealistically low
values for the electric field. Since the Debye length and inter-particle distance seem to be in agreement with other measurements, the unrealistically high negative charge 
implies that the value obtained for $k$ is too high, even though the value determined from the particle
trajectories seems to be rather precise. 

Since we did not include the ion drag in the vertical force balance, the radius of curvature of the cutout in the cover-plate
on top of the powered electrode determined from the value of $k_0$ obtained without additional heating is given by

\begin{equation}
R_C = \frac{m_Dg}{k_0}.
\end{equation}

\noindent However, if the ion drag does play a significant role, the radius of curvature would be given by

\begin{equation}
R_C = \frac{(m_Dg+F_{id})}{k_0},
\end{equation}

\noindent resulting in a larger radius of curvature, since the ion drag acts downwards and adds to gravity. The resulting
change in $k$ due to the thermophoretic force $F_{th}$ is then given by

\begin{equation}
\Delta k = k_{\Delta T} - k_0 =  \frac{-F_{th}}{R_C},
\end{equation}

\noindent which would be smaller due to the larger value of $R_C$. This would result in smaller charges for a calculated
${\lambda}_D$, which follows from equation \ref{eq:charge}. In order to have a significant change in the charge, the ion drag
would need at a minimum to be of the same order as gravity, which is not expected for particles of this size \cite{Hebner2002}.
Nonetheless, the ion drag seems to play an important role in the vertical force balance. 

Despite ignoring the ion drag in the analysis, our measurements show that the variation 
in the electric field can be probed on the sub-millimeter scale, and that the dust crystal is levitated 
on the plasma side of the Bohm point.

Of the three observables used in this determination, the number of particles turned out to be the hardest to obtain. This is due to a few factors.
First, in order to obtain high enough resolution to accurately measure the central inter-particle distance, the field of view
of the top-mounted
camera had to be reduced, so that the crystal did not fit completely in the frame. However, it was still possible to estimate what fraction
 of the crystal was missing and correct the number of particles. Secondly, in some cases the
horizontal laser sheath also illuminated particles below the top-layer, resulting in an overcount of the total number of particles. Finally, and most importantly, despite all our efforts, more and more
dust particles were lost from the top layer when we increased the temperature on the lower electrode. At this point, the initially
single-layer crystal seemed to split into additional layers. This was also observed with the side-view camera. As a result, the
equation of state, which strongly depends on the single-layer structure of the crystal, may no longer have been valid.

An interesting observation, which we could not find in the literature, was that a measurable change in the natural DC bias on the
lower electrode occured.
This bias became less negative for electrode temperatures above room-temperature, changing from -13 V when the electrode was at
room-temperature to roughly -10.5 V at +60${}^{\circ}$ C. When the electrode was cooled below room-temperature, we did not observe
the opposite effect. Since a decrease in the negative DC bias causes dust particles to levitate closer to the lower
electrode, negating the effect of the applied temperature, we maintained the DC bias at -13 V for all temperatures using an
external power supply. This insured that no changes in the DC bias could occur, so that the DC bias would not play a role in the levitation of the dust. 
Even so, very small changes in the bias can
not be excluded, and the question arises whether or not this might also have played a role in past experiments employing
thermophoresis.

\begin{figure}[!ht]
\centering
\includegraphics[width=2.5in]{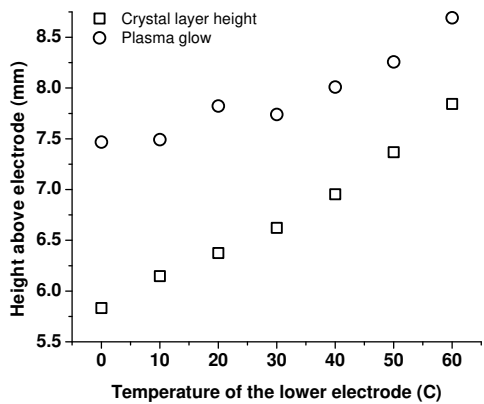}
\caption{The levitation height shown together with the height of the lower boundary of the plasma glow, measured from
side-view images taken at various temperatures. The glow appears to be moving together with the dust for higher
temperatures.}
\label{part_glow_height}
\end{figure}

Figure \ref{part_glow_height} shows the obtained crystal levitation height together with the lower boundary of the plasma
glow for different bottom electrode temperatures, estimated using images obtained with the side-view camera. We see that at higher temperatures, 
the plasma glow seems to move up together with the dust crystal. Even though dust particles have been shown to have a significant effect on
plasma emission in a dusty argon plasma \cite{Hubner2009}, we cannot at this point say whether this is due to the influence of the
dust crystal on the plasma, or due to the observed changes in the DC bias. This effect should be more carefully investigated in future
work.

\section*{Acknowledgment}

This research was made possible by NSF grant PHY-0648869 and NSF CAREER grant PHY-0847127.

\begin{IEEEbiography}[{\includegraphics[width=1in,height=1.25in,clip,keepaspectratio]{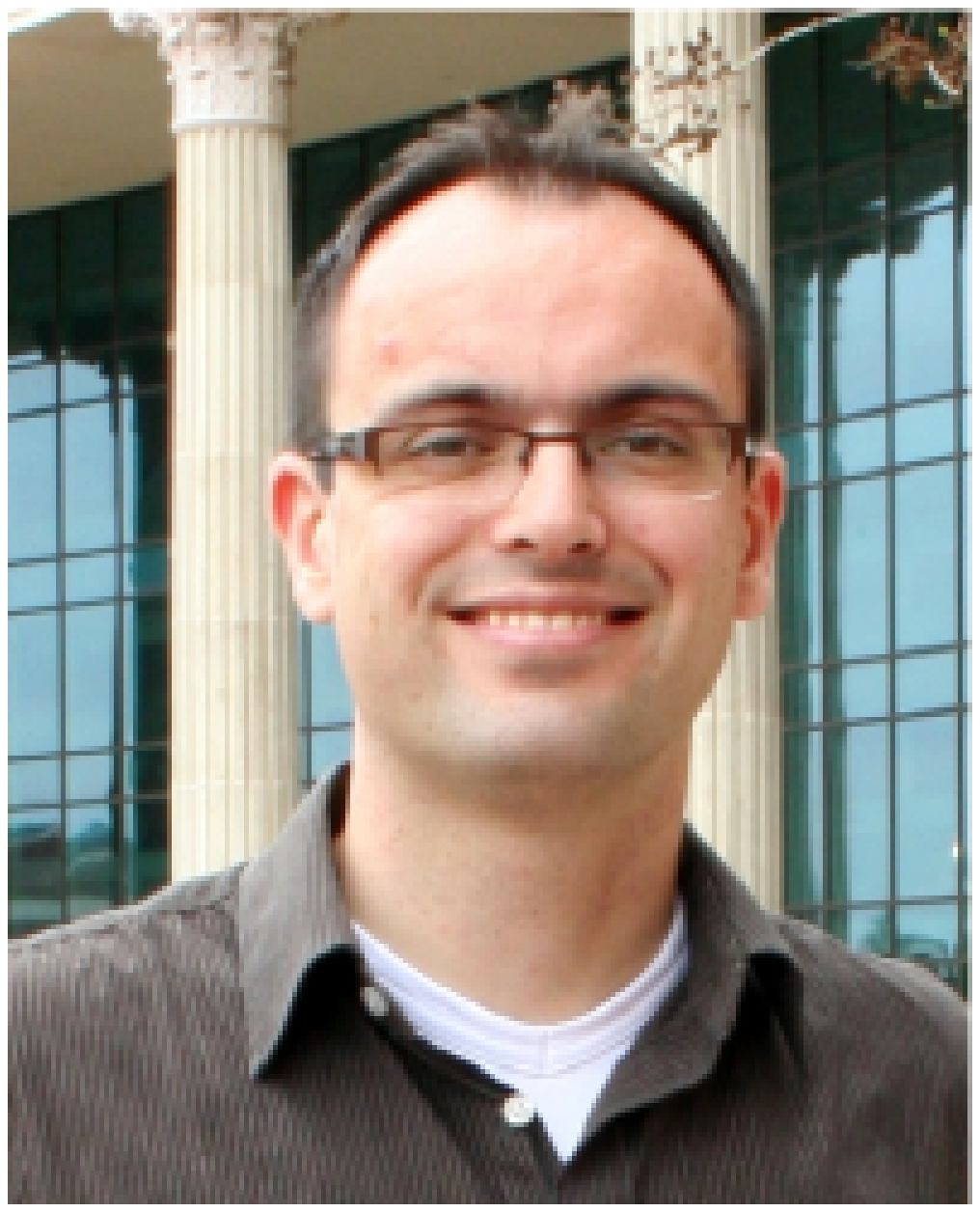}}]{Victor Land} was born in Petten, the Netherlands in 1979 and received his MSc in general astrophysics at Utrecht University in
the Netherlands in 2003, and his PhD at the
FOM-Institute for Plasma Physics 'Rijnhuizen', in the Netherlands in 2007. He is currently a post-doctorate research
associate at the Center for Astrophysics, Space Physics and Engineering Research at Baylor University, in Waco, Texas, where
he works on particle transport, charging, and coagulation in dusty plasma. 
\end{IEEEbiography}

\begin{IEEEbiography}[{\includegraphics[width=1in,height=1.25in,clip,keepaspectratio]{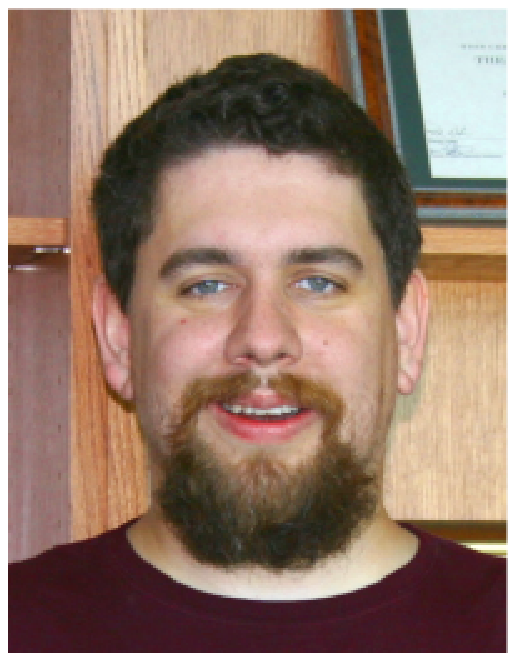}}]{Bernard Smith} was born in Rapid City, 
South Dakota, in 1975.  He received the B.S. degree in physics and math from Baylor University, Waco, Texas, in 1998, the M.S. degree 
in physics from Baylor University in 2003, and the Ph.D. degree in physics from Baylor University in 2005.
His current research interests include complex plasma physics and physics education.
\end{IEEEbiography}

\begin{IEEEbiography}[{\includegraphics[width=1in,height=1.25in,clip,keepaspectratio]{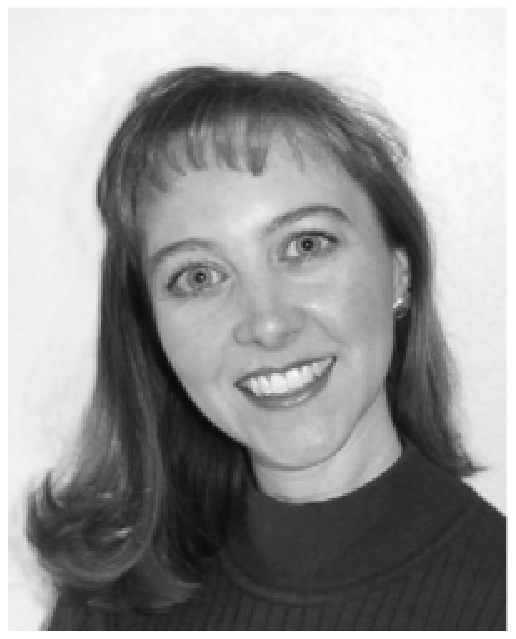}}]{Lorin Matthews}
was born in Paris, TX in 1972. She
received the B.S. and the Ph.D. degrees in physics from
Baylor University in Waco, TX, in 1994 and 1998,
respectively.
She is currently an Assistant Professor in the Physics
Department at Baylor University. Previously, she worked
at Raytheon Aircraft Integration Systems where she was
the Lead Vibroacoustics Engineer on NASA's SOFIA
(Stratospheric Observatory for Infrared Astronomy) project.
\end{IEEEbiography}

\begin{IEEEbiography}[{\includegraphics[width=1in,height=1.25in,clip,keepaspectratio]{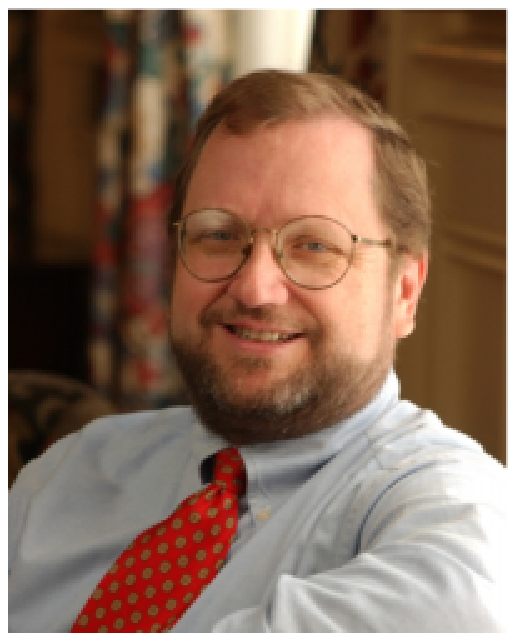}}]{Truell Hyde}
was born in Lubbock, Texas in 1956. He
received the B.S. in physics and mathematics from
Southern Nazarene University in l978 and the Ph.D. in
theoretical physics from Baylor University in 1988.
He is currently at Baylor University where he is the
Director of the Center for Astrophysics, Space Physics \&
Engineering Research (CASPER), a Professor of physics
and the Vice Provost for Research for the University. His
research interests include space physics, shock physics and waves and
nonlinear phenomena in complex (dusty) plasmas.
\end{IEEEbiography}

\end{document}